\RequirePackage{fix-cm}
\documentclass[smallcondensed]{svjour3}     
\smartqed  
\usepackage{graphicx}
\usepackage{amsmath}
\usepackage{graphicx}\usepackage{txfonts}
\usepackage{hyperref}
\usepackage{natbib}

\begin{document}

\title{The planar two-body problem for spheroids and disks 
}


\author{Margrethe Wold$^{*}$ \and
        John T. Conway 
}


\institute{
  Department of Engineering Sciences, University of Agder, Jon Lilletuns vei 9, N-4879 Grimstad\\
              Tel.: +47-37233476\\
              \email{margrethe.wold@uia.no}           
}

\date{Received: date / Accepted: date}

\maketitle

\begin{abstract}
  We outline a new method suggested by \citeauthor{conway16} (\citeyear{conway16})
  for solving the two-body problem for solid bodies of spheroidal or
  ellipsoidal shape. The method is based on integrating the gravitational potential of one
  body over the surface of the other body. When the gravitational potential can be analytically
  expressed (as for spheroids or ellipsoids), the gravitational force and mutual gravitational potential
  can be formulated as a surface integral instead of a volume integral, and solved numerically.
  If the two bodies are infinitely thin disks, the surface integral has an analytical solution.
  The method is exact as the force and mutual potential appear in closed-form expressions,
  and does not involve series expansions with subsequent truncation errors.
  In order to test the method, we solve the equations of motion in an inertial frame, and
  run simulations with two spheroids and two infinitely thin disks,
  restricted to torque-free planar motion. The resulting trajectories display precession
  patterns typical for non-Keplerian potentials. We follow 
  the conservation of energy and orbital angular momentum, and also investigate 
  how the spheroid model approaches the two cases where the surface integral can be solved
  analytically, i.e.\ for point masses and infinitely thin disks.
     
\end{abstract}


\section{Introduction}
\label{section:intro}

In celestial mechanics, a classical problem is to model the dynamics of two
rigid, extended bodies under mutual gravitational attraction. In the most
general case, the bodies have arbitrary shapes and can have both
translational and rotational motion,
yielding twelve degrees of freedom. Naturally, to model such a system is 
computationally expensive, and simplifications and approximations are
commonly made.

During the last 20 years, there has been renewed interest in the extended
two-body problem in astronomy as binary asteroids have been discovered
and studied in detail (e.g.\ \citeauthor{margot02} \citeyear{margot02}). The two components of a
binary asteroid can be
physically close and have irregular shapes, and their translational
and rotational motion is coupled through energy and torque transfer. 
To describe the dynamics of such a system requires the full two-body problem to be solved.

The mutual gravitational potential $U$ between two extended bodies A and B
of arbitrary shape is a six-dimensional integral as it involves volume
integration over each body: 
\begin{equation}
U=-G\int_{A}\int_{B}\frac{dm_{A}dm_{B}}{r},  \label{eq:eqn1}
\end{equation}
with $r$ being the distance between mass elements $dm_{A}$ and $dm_{B}$.
In the literature, a number of approaches are described to compute the mutual
potential, largely depending on the application.
It is customary to assume constant density and convert the integral to 
an integral of the scalar potential of one body over the volume of the
other. The integrand is expanded in terms of Legendre polynomials or
spherical harmonics, and higher order terms are neglected depending on the
required accuracy (e.g.\ \citeauthor{borderies78} \citeyear{borderies78};
\citeauthor{hartmann94} \citeyear{hartmann94}).
Another common approach is to use Cartesian coordinates instead of spherical
coordinates, and convert the integrals over mass elements to inertia
integrals (\citeauthor{maciejewski95} \citeyear{maciejewski95};
\citeauthor{ashenberg07} \citeyear{ashenberg07};
\citeauthor{scheeres09} \citeyear{scheeres09};
\citeauthor{bl09} \citeyear{bl09};
\citeauthor{jacobsonscheeres2011} \citeyear{jacobsonscheeres2011};
\citeauthor{hou2017} \citeyear{hou2017}).
This approach is motivated especially for non-spherical shapes and artificial satellites.

By using spherical harmonics and/or inertia integrals, the gravitational potential 
will always be an approximation because higher order terms in the series are neglected.
Truncation errors can become significant, especially close to the
surface of the bodies, or can accumulate in a manner that will cause problems
when describing long-term dynamical behaviour.
Furthermore, for spherical harmonics, the radius of convergence for the
approximating series is a circumscribing
sphere around each body, and within this sphere the series does not converge
(mitigated to some degree by using
spheroidal or ellipsoidal harmonics
(\citeauthor{jekeli88} \citeyear{jekeli88};
\citeauthor{gb01} \citeyear{gb01};
\citeauthor{fukushima14} \citeyear{fukushima14};
\citeauthor{rb16} \citeyear{rb16})).

Asteroid shapes can also be represented by polyhedra,
and several works exist where
binaries are modeled by two polyhedra (e.g.\
\citeauthor{ws05} \citeyear{ws05};
\citeauthor{fs06} \citeyear{fs06};
\citeauthor{hs13} \citeyear{hs13}).
However, in these cases the mutual potential is also
expressed as an infinite series where truncations become necessary for
practical purposes. A somewhat different approach is presented by
\citeauthor{shi17} (\citeyear{shi17}) where one asteroid is modeled
as a homogeneous polyhedron with a closed-form gravitational
potential thus avoiding truncation, and the other 
body is modeled with an 
arbitrary mass distribution, and the mutual potential is expressed by 
inertia integrals truncated at second order.

The methods listed above have in common that they all involve series expansions to
describe the mutual potential and the dynamics of two extended rigid bodies.
In this paper we utilize the approach used by 
\citeauthor{conway16} (\citeyear{conway16})
who explored the use of integral theorems to express the force, torque and
mutual potential between two gravitating bodies as two-dimensional surface integrals over
one or both bodies. 
We test the application of two of the surface integral equations in a dynamical context for two
extended bodies under mutual gravitational interaction. To our knowledge, this scheme has not
been tried for two extended bodies before. We have chosen to 
concentrate on planar motion of spheroids and thin disks as they have
well-known analytical expressions
for the gravitational potential (\citeauthor{macmillan} \citeyear{macmillan}),
and as long as this is the case, this method also avoids series expansion.
The equations of motion are solved using a Runge-Kutta integrator,
and we show example orbits and address changes
in orbital angular momentum and total energy as a probe of numerical accuracy.

The paper is organized as follows:
In Sect.~\ref{section:sec0} we introduce the integral equations based on Conway's
(\citeyear{conway16}) description and in Sect.~\ref{section:sec1} we outline a surface
integration method that can be
applied to spheroids and/or ellipsoids, and use it to compute the force and mutual potential
for the case of two spheroids in Sect.~\ref{section:sec2}.
In Sect.~\ref{section:sec3} we outline how to find the force and mutual potential between two
coplanar thin disks using an analytical expression with complete elliptical integrals. 
We thereafter run simulations with two spheroids and two disks, and present the results 
in Sect.~\ref{section:sec4}, where we also investigate the conservation of
energy and angular momentum. In
Sect.~\ref{section:sec3a} we compare pairs of spheroids having extreme shapes
(nearly spherical and significantly flattened) to analytical solutions
for point masses/spheres and infinitely thin disks. 
We discuss the results and some applications, as well as prospects for
future work in Sect.~\ref{section:sec5}.

\section{Surface integral equations for force, torque and mutual potential}
\label{section:sec0}

For the general extended two-body problem, three integrals are needed for the force,
torque and mutual potential $U$. The force $\mathbf{F}$ on a body of constant density
$\rho$ is given by the
well-known surface integral 
\begin{equation}
\mathbf{F} = \rho \iint\nolimits_{S}\Phi \left( \mathbf{r}\right)\, \mathrm{d}\mathbf{S}
\label{eq:eqn2}
\end{equation}%
where $\mathrm{d}\mathbf{S}$ is a vector perpendicular to the surface $S$ of the
body and $\mathbf{r}$ is a position vector to a point on the surface of the body being
integrated over.
This equation is immediately obtained by applying Gauss's theorem
to the volume integral for the force $\mathbf{F}$:
\begin{equation}
\mathbf{F} = \rho \iiint\nolimits_{V} \mathbf{g}(\mathbf{r})\, {\rm d}V
\end{equation}
where $\mathbf{g}(\mathbf{r}) = \nabla\Phi(\mathbf{r})$ is the gravitational field acting
on the the body at any point $\mathbf{r}$. The corresponding volume integral for the torque
$\mathbf{M}$ is
\begin{equation}
\mathbf{M} = \rho\iiint\nolimits_{V} \mathbf{r} \times \mathbf{g}(\mathbf{r})\, \mathrm{d}V. 
\end{equation}
\citeauthor{conway16} (\citeyear{conway16}) introduced two alternative
vector potentials
$\mathbf{V}_1(\mathbf{r})$ and $\mathbf{V}_2(\mathbf{r})$ for the vector
$\mathbf{r} \times \mathbf{g}(\mathbf{r})$, with
$\nabla \times \mathbf{V}_1 = \nabla \times \mathbf{V}_2 = \mathbf{r}
\times \mathbf{g}(\mathbf{r})$
where
\begin{equation}
  \mathbf{V}_1 = -\mathbf{r}\Phi(\mathbf{r})
  \end{equation}
and
\begin{equation}
  \mathbf{V}_2 = \frac{1}{2}\lvert \mathbf{r} \rvert^2 \mathbf{g}(\mathbf{r}).
\end{equation}
Substituting these vector potentials into the well-known theorem
\begin{equation}
\iiint\nolimits_{V} \nabla \times \mathbf{V}(\mathbf{r})\, \mathrm{d}V = \iiint\nolimits_{S} \mathbf{n} \times \mathbf{V}(\mathbf{r})\, \mathrm{d}S
\end{equation}
gives two alternative surface integrals for $\mathbf{M}$:
\begin{equation}
\mathbf{M} = -\rho\iint\nolimits_S \Phi(\mathbf{r})\, \mathbf{n}\times \mathbf{r}\, \mathrm{d}S 
\label{eq:torque1}
\end{equation}
\begin{equation}
\mathbf{M} = \frac{\rho}{2}\iint\nolimits_S \lvert \mathbf{r} \rvert^2\, \mathbf{n}\times\mathbf{g}(\mathbf{r})\, \mathrm{d}S. 
\label{eq:torque2}
\end{equation}
A third vector potential for $\mathbf{r}\times \mathbf{g}(\mathbf{r})$ can be defined as
$\mathbf{V}_3(\mathbf{r}) = (\mathbf{V}_1 + \mathbf{V}_2)/2$,
which has the property
\begin{equation}
  \nabla \cdot \mathbf{V}_3 = -\frac{3}{2} \Phi(\mathbf{r})
\end{equation}
The volume integral for the mutual potential is 
\begin{equation}
U = \rho \iiint\nolimits_{V}\Phi(\mathbf{r})\, {\rm d}V,
\end{equation}
and employing $\nabla \cdot \mathbf{V}_3(\mathbf{r})$ in the divergence theorem immediately gives a
surface integral for $U$ as
\begin{equation}
U=\frac{\rho}{3}\iint_{S}\left[ \mathbf{r}\Phi\left( \mathbf{r}
\right) -\frac{1}{2}|\mathbf{r}|^{2}\mathbf{g}\left( \mathbf{r}\right) 
\right] \cdot \,\mathrm{d}\mathbf{S}.  \label{eq:pot_energy}
\end{equation}

Eqs.~\ref{eq:eqn2}, \ref{eq:torque1}/\ref{eq:torque2} and \ref{eq:pot_energy} are
surface integral equations for force, torque and mutual potential, hence the integral dimension
is reduced by one compared to the more common volume integrals. 
For the planar motion of spheroids considered here, the torque integrals are not needed, as no
distribution of matter can induce a torque about an axis of rotational symmetry of a body.
In a future paper we will implement torques 
(Ho et al.\ in prep.).

In this paper, we test the application of Eqs.~\ref{eq:eqn2} and \ref{eq:pot_energy} in 
a dynamical context for two extended bodies under mutual gravitational interaction. In the next
section, we outline our method of surface integration for spheroids/ellipsoids.


\section{Force and mutual gravitational potential: surface integration method}

\label{section:sec1}

\begin{figure}
\includegraphics[width=1.0\textwidth]{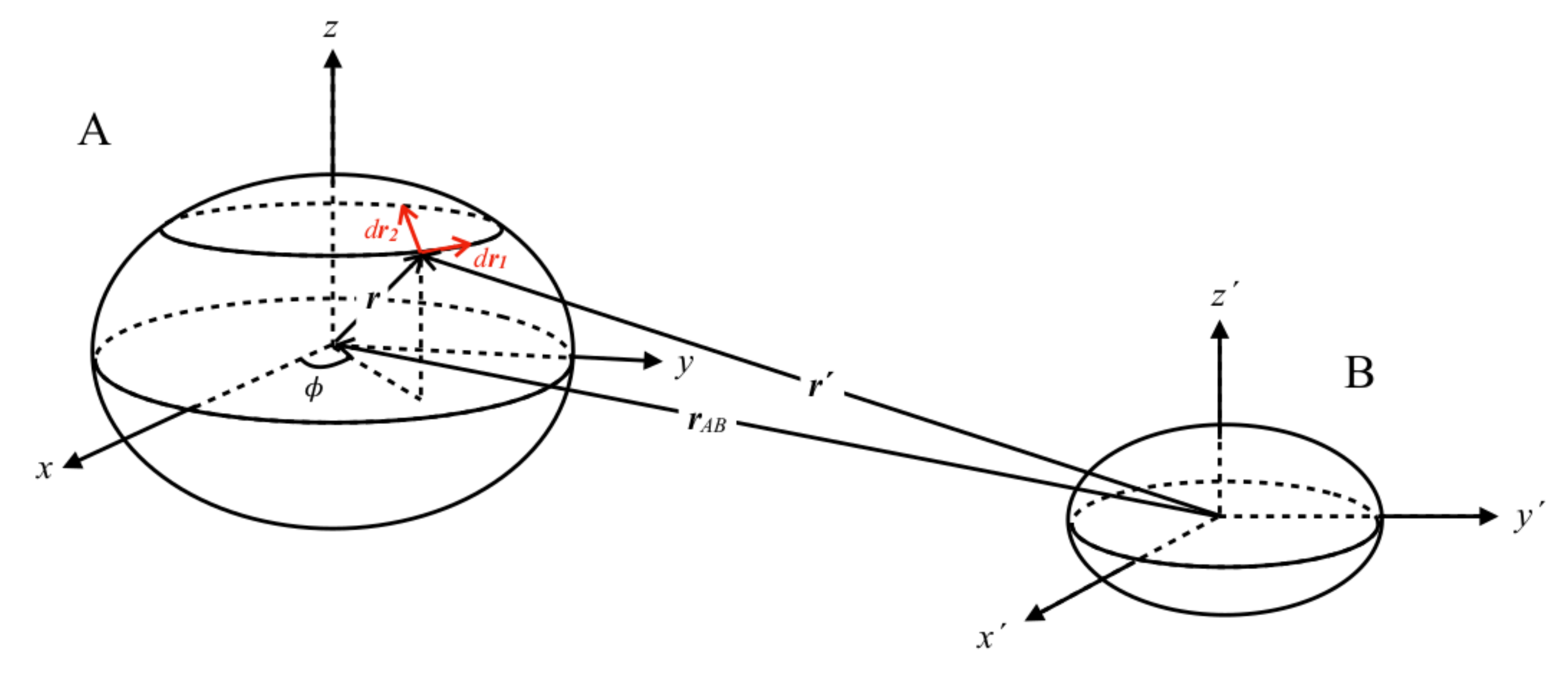}
\caption{Illustration showing our use of variables and coordinate systems. A plane parallel to the $xy$-plane of A intersects
  the ellipsoid and forms an ellipse traced out by azimuthal angle $\phi$ and radius vector $\mathbf{r}$.
  We integrate over the surface of A, and the area element
  $\mathrm{d}S$
is spanned by the two vectors $\mathrm{d}\mathbf{r_1}$ and $\mathrm{d}\mathbf{r_2}$} 
\label{fig:fig1}
\end{figure}

Assume we have two homogeneous, rigid bodies A and B with constant mass densities
$\rho_A$ and $\rho_B$ and that their respective body-fixed coordinate systems
are $(x,y,z)$ and $(x',y',z')$, see Fig.\ref{fig:fig1}. 
Body B has gravitational potential $\Phi_{B}(\mathbf{r})$ where $\mathbf{r}=[x,y,z]$
is a position vector
in A's body-fixed frame (for now, we choose to express B's gravitational potential in A's
coordinate system). The force on body A from body B is given by the surface integral:
\begin{equation}
\mathbf{F}_A = \rho_A \iint_{S_A} \Phi_B(\mathbf{r})\, \mathrm{d}\mathbf{S},
\label{Eqn_surfint}
\end{equation}
where $\mathrm{d}\mathbf{S}$ is an outwardly pointing vector normal to the bounding surface $S_A$ of
A. Assume that the potential $\Phi_{B}(\mathbf{r})$ is known and can be evaluated at all positions on A's surface.

The next step is to do the surface integration over body A. We wish to do the surface integration over A in
A's body-fixed frame, and use azimuthal angle $\phi$ and $z$-coordinate as integration variables,
see Fig.\ref{fig:fig1}.
We assume that A has the shape of an ellipsoid with semi-axes $a>b>c$. The surface of A is therefore described by: 
\begin{equation}
\left( \frac{x}{a}\right) ^{2}+\left( \frac{y}{b}\right) ^{2}+\left( \frac{z}{c}\right) ^{2}=1\text{.}  \label{Eqn1}
\end{equation}
The cross section of a plane parallel to the $xy$-plane and the ellipsoid surface is an ellipse with 
equation
\begin{equation}
\left( \frac{x}{\alpha}\right) ^{2}+\left( \frac{y}{\beta}\right)
^{2}=1  \label{Eqn2}
\end{equation}
where $\alpha$ and $\beta$ are: 
\begin{equation}
\alpha =\frac{a}{c}\sqrt{c^{2}-z^{2}}  \label{Eqn3}
\end{equation}
\begin{equation}
\beta =\frac{b}{c}\sqrt{c^{2}-z^{2}}  \label{Eqn4}
\end{equation}
(obtained by substituting the equation for a plane parallel to the $xy$-plane into the ellipsoid equation).
Hence, lines of constant $z$ on the ellipsoid surface can be expressed by Eq.~\ref{Eqn2}, see also Fig.~\ref{fig:fig1}. We parametrise these ellipses in terms of azimuthal angle $\phi$ in the usual manner as:
\begin{equation}
x=\alpha \cos \phi  \label{Eqn5}
\end{equation}
\begin{equation}
y=\beta\sin \phi \text{.}  \label{Eqn6}
\end{equation}
For finding the area element vector $\mathrm{d}\mathbf{S}$, we start with small displacements $\mathrm{d}\mathbf{r}_{1}$
and $\mathrm{d}\mathbf{r}_{2}$ on the ellipsoid surface along constant $z$ and constant $\phi$, respectively. We refer to Fig.~\ref{fig:fig1}
in the following.
The small displacement $\mathrm{d}\mathbf{r}_{1}$ along a constant $z$ is given by 
\begin{equation}
\mathrm{d}\mathbf{r}_{1} =
\begin{bmatrix}
  \mathrm{d}x \\
  \mathrm{d}y \\
  0 \\
\end{bmatrix}
  = 
    \begin{bmatrix}
-\frac{a\sqrt{c^{2}-z^{2}}}{c}\sin \phi \\ 
\frac{b\sqrt{c^{2}-z^{2}}}{c}\cos \phi \\ 
0
\end{bmatrix}
\mathrm{d}\phi \text{,}  \label{Eqn7}
\end{equation}
where we have differentiated Eq.~\ref{Eqn5} and \ref{Eqn6}.
A corresponding displacement $\mathrm{d}\mathbf{r}_{2}$ along a line of
constant $\phi$ is given by 
\begin{equation}
  \mathrm{d}\mathbf{r}_{2}=%
\begin{bmatrix}
  \mathrm{d}x \\
  \mathrm{d}y \\
  \mathrm{d}z \\
\end{bmatrix}
= 
\begin{bmatrix}
-\frac{az}{c\sqrt{c^{2}-z^{2}}}\cos \phi \\ 
-\frac{bz}{c\sqrt{c^{2}-z^{2}}}\sin \phi \\ 
1
\end{bmatrix}
\mathrm{d}z\text{.}  \label{Eqn8}
\end{equation}
In general the line elements $\mathrm{d}\mathbf{r}_{1}$ and $\mathrm{d}%
\mathbf{r}_{2}$ are parallel to the sides of a parallelogram, and for a
triaxial ellipsoid they are orthogonal only along certain symmetric curves.
For the surface of a spheroid and a sphere, the line elements are orthogonal
everywhere.

The area element $\mathrm{d}\mathbf{S}$ is evaluated as the cross product between $\mathrm{d}\mathbf{r_1}$ and
$\mathrm{d}\mathbf{r_2}$:
\begin{equation}
\mathrm{d}\mathbf{S=}\, \mathrm{d}\mathbf{r}_{1} \times \mathrm{d}\mathbf{r}_{2} =
\begin{bmatrix}
\frac{b\sqrt{c^{2}-z^{2}}}{c}\cos \phi \\ 
\frac{a\sqrt{c^{2}-z^{2}}}{c}\sin \phi \\ 
\frac{ab}{c^{2}}z
\end{bmatrix}
\mathrm{d}\phi\, \mathrm{d}z\text{.}  \label{Eqn12}
\end{equation}
With $\mathrm{d}\mathbf{S}$ from Eq.~\ref{Eqn12} we can evaluate the
surface integral for the force in Eq.~\ref{Eqn_surfint} if the gravitational potential of
body B is known. The force on body A can thus be found, and by Newton's
third law also the force on body B:

\begin{equation}
  \mathbf{F}_A = -\mathbf{F}_B = \rho_A \int_{-c}^{c}\int_{0}^{2\pi} \Phi_B(\mathbf{r})\,
    \begin{bmatrix}
\frac{b\sqrt{c^{2}-z^{2}}}{c}\cos \phi \\ 
\frac{a\sqrt{c^{2}-z^{2}}}{c}\sin \phi \\ 
\frac{ab}{c^{2}}z
    \end{bmatrix}
    \mathrm{d}\phi\, \mathrm{d}z\text{.}  \label{eq:force_eqn}
\end{equation}
If the gravitational potential of B is a function that can be evaluated at each position
$\mathbf{r}$ on A's surface, the force can be computed with this double integral.

For the potential energy $U$ given by Eq.~\ref{eq:pot_energy} we need to evaluate the two terms
$\mathbf{r} \Phi_{B}(\mathbf{r}) \cdot \mathrm{d} \mathbf{S}$ and
$\frac{1}{2}|\mathbf{r}|^2 \mathbf{g}_{B} (\mathbf{r}) \cdot \mathrm{d}\mathbf{S}$.
The first term is straightforward, and to compute the
second term, the gravitational field is found from the gradient of the potential, 
$\mathbf{g}_{B}(\mathbf{r}) = \nabla \Phi_{B}(\mathbf{r})$. In cases where the potential of B can be expressed analytically, exact values of $\Phi_{B}$ and $\mathbf{g}_{B}$ can be computed
at each point $\mathbf{r}$ on A's surface during the integration.

\subsection{Tests for two spheroids}

\label{section:sec2}

In this paper, we wish to test the use of Eqs.~\ref{eq:pot_energy} and \ref{eq:force_eqn}
in a relatively simple case where both bodies are spheroids with $a=b$.
This allows us to use the integration scheme outlined above, and to utilize
the known analytic expressions
for the gravitational potential of oblate ($a>c$) spheroids (prolate spheroids with $a<c$ were also tested). 
In order to clarify our use of symbols, we add subscripts to the spheroid
shape parameters so that $a_A (=b_A)$ becomes the
equatorial radius of spheroid A, and $c_A$ the distance from the centroid of
A to its pole along the symmetry axis ($z$-axis). Similarly for spheroid B.
For an oblate spheroid B, the gravitational potential at an exterior field point is given by
the following analytic expression (\citeauthor{macmillan} \citeyear{macmillan}): 
\begin{multline}
\Phi_B\left(x',y',z'\right) = \frac{2\pi\rho_B a_B^2 c_B}{\sqrt{a_B^2 - c_B^2}}%
\left(1-\frac{x'^{2}+y'^{2}-2z'^{2}}{2(a_B^2-c_B^2)}\right)\sin^{-1}\sqrt{\frac{a_B^2-c_B^2}{a_B^2+\kappa}} \, + \\
\frac{\pi\rho_B a_B^2 c_B \sqrt{c_B^2+\kappa}}{a_B^2-c_B^2} \frac{x'^{2}+y'^{2}}{a_B^2+\kappa} - 
\frac{\pi\rho_B a_B^2 c_B}{a_B^2-c_B^2}\frac{2z'^{2}}{\sqrt{c_B^2+\kappa}},
\label{Eqn_phi_spheroid}
\end{multline}
where $\kappa$ is defined by the equation: 
\begin{equation}
\frac{x'^{2}+y'^{2}}{a_B^2+\kappa} + \frac{z'^{2}}{c_B^2+\kappa} = 1
\label{Eqn_kappa}
\end{equation}
(see \citeauthor{macmillan} (\citeyear{macmillan}) for the corresponding equation for a prolate
spheroid). As mentioned above, $\Phi_{B}(\mathbf{r})$ appearing in the double integral is the 
gravitational potential of B expressed in A's coordinate system. However, MacMillan's expression above
uses $\mathbf{r'}=[x',y',z']$ which is the position vector of a field point in the body-fixed frame of B with
B's centroid at the origin. During the integration, when calling the functional form given in
Eq.~\ref{Eqn_phi_spheroid}, we therefore replace $\mathbf{r'}$ with $\mathbf{r}+\mathbf{r}_{AB}$,
where $\mathbf{r}_{AB}$ is the vector from B's to A's centroid (see Fig.~\ref{fig:fig1}).

For evaluating the integral for $U$ in Eq.~\ref{eq:pot_energy},
the gravitational field has to be evaluated as
$\mathrm{\mathbf{g}_B(\mathbf{r})} = \nabla \Phi(\mathbf{r})$.
Differentiating Eq.~\ref{Eqn_phi_spheroid} yields:
\begin{equation}
  \mathbf{g}_B(\mathbf{r'}) =
2\pi \rho_B a_B^2c_B^2 
  \begin{bmatrix}
    x'(f_1-f_2) \\
    y'(f_1-f_2)\\
    2z'(f_2-f_3) \\
  \end{bmatrix}
\end{equation}
where
\begin{equation}
f_1 = \frac{c_B^2 + \kappa}{(a_B^2 - c_B^2)(a_B^2 + \kappa)},
\end{equation}
\begin{equation}
f_2 = \sin^{-1}\sqrt{\frac{a_B^2-c_B^2}{a_B^2+\kappa^2}} \frac{1}{(a_B^2-c_B^2)^{3/2}}
\end{equation}
and
\begin{equation}
f_3 = \frac{1}{(a_B^2 - c_B^2)\sqrt{c_B^2+\kappa}}
\end{equation}
Again, when calling the function $\mathbf{g}_{B}$ in the surface integral, the position vector changes from $\mathbf{r}'$
to $\mathbf{r}$. 

By using the analytical expression for $\Phi_B(\mathbf{r'})$ as well as the
surface element $\mathrm{d}\mathbf{S}$ defined in Eq.~\ref{Eqn12} in the two integral equations,
we obtain expressions for the force and mutual potential 
that are {\em exact} in the sense that they do not involve approximations that normally follow when
gravitational potentials are written as truncated series. With this method,
the expressions for the force
and potential energy are integral equations and therefore not analytical.  
The integrals have to be solved numerically, hence with ``exact'' we mean exact to within the limits
of numerical integration. Here, we solve the surface integrals numerically by using
the Gaussian quadrature integration scheme for double integrals implemented in Python
({\tt integrate.dlbquad}) with the relative and absolute tolerance set to the
default value of $1.49\times10^{-8}$.

As far as the authors are aware, the surface integrals for gravitational force and mutual potential energy
between two non-spherical bodies 
can be analytically expressed in just one case, which is that of two coplanar, thin, non-coaxial disks. We
investigate this case in the next section. 

\subsection{Tests for two coplanar disks}

\label{section:sec3}

\citeauthor{conway16} (\citeyear{conway16}) has also derived the gravitational force between two rigid,
coplanar, non-coaxial, infinitely thin disks with masses $m_{A}$ and $m_{B}$
and radii $R_{A}$ and $R_{B}$. As opposed to the case with two spheroids, the
surface integrals for force and mutual gravitational potential have analytic forms.
The expression is in closed form in terms of complete elliptic integrals, and we also 
wish to test this solution. The force on one disk from the other is given by: 
\begin{equation}
F(R_{A},R_{B},r_{AB})=-\frac{Gm_{A}m_{B}}{r_{AB}^{2}}f(R_{A},R_{B},r_{AB}),
\end{equation}%
where $r_{AB}$ is the distance between the axes of the disks, and $%
f(R_{A},R_{B},r_{AB})$ is a dimensionless shape factor representing the deviation
from an inverse square law: 
\begin{multline}
f\left( R_{A},R_{B},r_{AB}\right) = 
\frac{8}{\pi ^{2}}\Big\{\left[ \frac{\mathbf{K}(k_{+})}{1-k_{+}^{2}}-\frac{%
\mathbf{K}(k_{+})-\mathbf{E}(k_{+})}{k_{+}^{2}(1-k_{+}^{2})}\right] \left[ 
\frac{\mathbf{K}(k_{-})}{1-k_{-}^{2}}-\frac{\mathbf{K}(k_{-})-\mathbf{E}%
(k_{-})}{k_{-}^{2}(1-k_{-}^{2})}\right]  \\
-\frac{1}{3}\left[ \frac{\mathbf{K}(k_{+})}{1-k_{+}^{2}}+\frac{(1-2k_{+}^{2})(\mathbf{K}(k_{+})-\mathbf{E}(k_{+}))}{k_{+}^{2}(1-k_{+}^{2})}\right] 
\left[ \frac{\mathbf{K}(k_{-})}{1-k_{-}^{2}}+\frac{(1-2k_{-}^{2})(\mathbf{K}(k_{-})-\mathbf{E}(k_{-}))}{k_{-}^{2}(1-k_{-}^{2})}\right] \Big\}.
\label{eq:diskforce}
\end{multline}%
Here $\mathbf{K}$ and $\mathbf{E}$ are the complete elliptic integrals of
first and second kind, respectively. The two factors $k_{+}$ and $k_{-}$
are given by: 
\begin{equation}
k_{\pm }^{2}=\frac{r_{AB}^{2}\pm (R_{A}^{2}-R_{B}^{2})-\sqrt{%
    (r_{AB}^{2}-R_{A}^{2}-R_{B}^{2})^{2}-4R_{A}^{2}R_{B}^{2}}}{2r_{AB}^{2}}.
\label{eq:k_def}
\end{equation}%
For the two disks, the mutual gravitational potential has the following analytic
expression (Eq. 151 and 152 by \citeauthor{conway16} (\citeyear{conway16})): 
\begin{equation}
U=\frac{Gm_{A}m_{B}}{r_{AB}}u\left( R_{A},R_{B},r_{AB}\right) 
\label{equation:Epot_disks}
\end{equation}%
where $u$ is another dimensionless shape factor: 
\begin{multline}
u\left( R_{A},R_{B},r_{AB}\right) =\frac{4}{\pi ^{2}}\mathbf{K}(k_{+})\mathbf{K}(k_{-}) 
+\frac{4R_{A}^{2}}{\pi ^{2}r_{AB}^{2}}\left( 2\frac{\mathbf{K}(k_{+})-\mathbf{E}%
(k_{+})}{k_{+}^{4}}-\frac{\mathbf{K}(k_{+})}{k_{+}^{2}}\right) \left( \frac{%
\mathbf{K}(k_{-})}{1-k_{-}^{2}}-\frac{2\mathbf{E}(k_{-})}{\left(
1-k_{-}^{2}\right) ^{2}}\right)  \\
+\frac{4R_{B}^{2}}{\pi ^{2}r_{AB}^{2}}\left( \frac{\mathbf{K}(k_{+})}{1-k_{+}^{2}}%
-\frac{2\mathbf{E}(k_{+})}{\left( 1-k_{+}^{2}\right) ^{2}}\right) \left( 2%
\frac{\mathbf{K}(k_{-})-\mathbf{E}(k_{-})}{k_{-}^{4}}-\frac{\mathbf{K}(k_{-})%
}{k_{-}^{2}}\right)  \\
+\frac{4R_{A}^{2}R_{B}^{2}}{9\pi ^{2}r_{AB}^{4}}\left( \frac{2\left(
1-2k_{+}^{2}\right) \left( \mathbf{K}\left( k_{+}\right) -\mathbf{E}\left(
k_{+}\right) \right) }{k_{+}^{4}\left( 1-k_{+}^{2}\right) ^{2}}-\frac{\left(
1-3k_{+}^{2}\right) \mathbf{K}(k_{+})}{k_{+}^{2}\left( 1-k_{+}^{2}\right)
^{2}}\right)  \\
\times \left( \frac{2\left( 1-2k_{-}^{2}\right) \left( \mathbf{K}(k_{-})-%
\mathbf{E}(k_{-})\right) }{k_{-}^{4}\left( 1-k_{-}^{2}\right) ^{2}}-\frac{%
\left( 1-3k_{-}^{2}\right) \mathbf{K}(k_{-})}{k_{-}^{2}\left(
1-k_{-}^{2}\right) ^{2}}\right) ,
\end{multline}%
and $\mathbf{K}$, $\mathbf{E}$ and $k_{+/-}$ have the same meaning as in
Eq.~\ref{eq:diskforce} and \ref{eq:k_def}.

It is straightforward to balance the gravitational and centrifugal forces for rotation of two such disks in circular coplanar orbits at constant distances from the centre of mass of the disks. This seems to be the only known truly analytical solution for the motion of two extended bodies when neither body has spherical symmetry.


\section{Results}

\label{section:sec4}

\subsection{Two coplanar spheroids}

Various tests were run with two spheroids of varying shapes and initial parameters. With the forces determined, the
equations of motion for the two bodies become a system of six first-order differential equations which
can be solved as a standard initial value problem. We solve the equations of motion in an
inertial frame using a fourth-order Runge-Kutta integrator (RK45) with a fixed time step.
There are two issues with this integrator: First, Runge-Kutta integrators are known to cause a drift
in energy, but as we have restricted ourselves to only shorter simulations covering a few
orbits, not much energy drift is expected to occur. 
Secondly, by using a fixed time step integration errors will
increase at periapsis, in particular for systems where the bodies are close.
However, since the simulations span only a few orbits, we have chosen a small time step which will
contribute toward minimizing integration errors. We include an error analysis in
Sect.~\ref{sec:erroranalysis} by following the conservation of energy and angular momentum
and find that the errors are acceptable for the demonstration purposes of this paper.

In the following,
we make the simplifying assumption that the two spheroids have a common
equatorial plane, i.e.\ they share a common $xy$-plane. Hence, there will be
no exchange of angular momentum between the two bodies and all torques vanish. The two
spheroids can be given arbitrary initial spins about their symmetry axes, but these spins will
remain constant throughout the motion.

We probe the conservation of total energy and total orbital angular momentum as
a function of time. The kinetic energy is readily computed from the velocities,
and the gravitational potential energy from Eq.~\ref{eq:pot_energy}. 
The magnitude of the total orbital angular momentum $J$ is  
\begin{equation}
J = \lvert \mathbf{R}_{A} \times m_{A}\mathbf{V}_{A} + \mathbf{R}_{B} \times
m_{B}\mathbf{V}_{B} \rvert,
\end{equation}
where $\mathbf{R}$ and $\mathbf{V}$ are positions and velocities of A
and B in the inertial frame (we denote all quantities relative to the
inertial frame with capital symbols).

We tested several cases of spheroid pairs with different shapes and
initial parameters, but show only one case with two identical, oblate
spheroids here. The parameters for the two spheroids are
given in Table~\ref{table:Table1}, and their tracjectories are shown in 
the top row of Fig.~\ref{fig:fig2}. The 
initial centroid-to-centroid distance (for the chosen initial velocity)
for the two bodies is rather small, corresponding to 7 length units.
Applying this, the
bodies reach a minimum distance of $r_{AB}\approx 2.55$ length units in the
simulation,
i.e.\ almost touching since they both have radii of $1.0$.
All quantities are kept unit-less.

\begin{table}
  \caption{Parameters for the two models discussed in the text,
    one with two identical spheroids ('2S') and one with two
    identical thin disks ('2D'). The mass of the bodies is 
    $m$, spheroid axes are $(a,c)$ and disk radii are
    $R$ (denoted $R_A$ and $R_B$ in Eqs.\ref{eq:diskforce}
    and \ref{equation:Epot_disks}). The last column is the time step
    used in the simulation. All quantities are unit-less}
\begin{tabular}{llll}
\hline\noalign{\smallskip}
Model  & Mass $m$ & Shape & $\Delta t$ \\ 
  \noalign{\smallskip}\hline\noalign{\smallskip}
  2S & $1.00$ & $(a,c) = (1,0.25)$ & $0.1$ \\
  2D & $1.00$ & $R = (1.00)$ & $0.01$ \\
\noalign{\smallskip}\hline
\end{tabular}
\label{table:Table1}
\end{table}

\begin{table}
  \caption{The characteristics of the $\Delta E$ and $\Delta J$ distributions 
    for the simulated cases, with $\Delta E$ and $\Delta J$
    being the difference in
    total energy and angular momentum between successive time steps.
The columns named 'slope' contain the slope of a straight line fit to
$\Delta E$ and $\Delta J$ versus time}
\label{table:errors}
\begin{tabular}{crrrr}
  \hline\noalign{\smallskip}
  Model & Mean $\Delta E$ & $\sigma$ & $\sigma/\sqrt{N}$ & Slope \\
  \noalign{\smallskip}\hline\noalign{\smallskip} 
2S & $-3\times10^{-11}$ & $3\times10^{-08}$ & $5\times10^{-10}$ & $-4\times10^{-10}$ \\
2D & $7\times10^{-09}$ & $6\times10^{-06}$ & $3\times10^{-08}$ & $-8\times10^{-07}$  \\
\noalign{\smallskip}\hline\noalign{\smallskip}
Model & Mean $\Delta J$ & $\sigma$ & $\sigma/\sqrt{N}$ & Slope \\
  \noalign{\smallskip}\hline\noalign{\smallskip}
2S & $1\times10^{-10}$ & $7\times10^{-08}$ & $1\times10^{-09}$ & $1\times10^{-09}$ \\  
2D & $-1\times10^{-18}$ & $2\times10^{-16}$ & $9\times10^{-19}$ & $-9\times10^{-17}$ \\
\noalign{\smallskip}\hline
\end{tabular}

\end{table}

The orbits of A and B around the system centre of mass (at the origin)
are confined between an inner radius, the pericentre
distance, and an outer radius, the apocentre distance. The precession
pattern is characterized by a radial and an azimuthal period, where
the radial period, 
$T_{r}$, is the time it takes for one of the bodies to travel from apocentre
to pericentre and back to apocentre, and the azimuthal period $T_{\psi }$ is
the period it takes for one of the bodies to travel $2\pi$ radians
(\citeauthor{btbook} \citeyear{btbook}).
The non-closed orbits shown in Fig.\ref{fig:fig2} have 
$T_{r}>T_{\psi }$, hence the bodies, upon passing the apocentre the second
time, have completed more than $2\pi $ radians around their common centre of
mass. The non-closed orbits are approximate ellipses where the major axis
precesses with an angle $\psi _{p}$ equal to the angle overshooting $2\pi $
radians for each radial period. For the case displayed in Fig.~\ref{fig:fig2},
the precession angle is $\psi _{p}=18^{\circ}$, and the precession rate
defined as $\Omega _{p}=\psi _{p}/T_{r}$ is $23^{\prime}$ per time
unit. It is positive, indicating that the major axis of the
ellipse rotates in the same direction as the bodies rotate around the centre
of mass. For two identical prolate spheroids with $(a,c)=(0.25,1.0)$ and
with the same masses and initial parameters as the two oblate ones, the
precession pattern rotates in the opposite direction with $\psi_{p}=-11^{\circ}$
($T_{r}<T_{\psi}$) and $\Omega_{p}=-14^{\prime}$
per time unit.

\subsection{Two thin, non-coaxial disks}
\label{section:disk_subsec}

\begin{figure}
\includegraphics[width=1.0\textwidth]{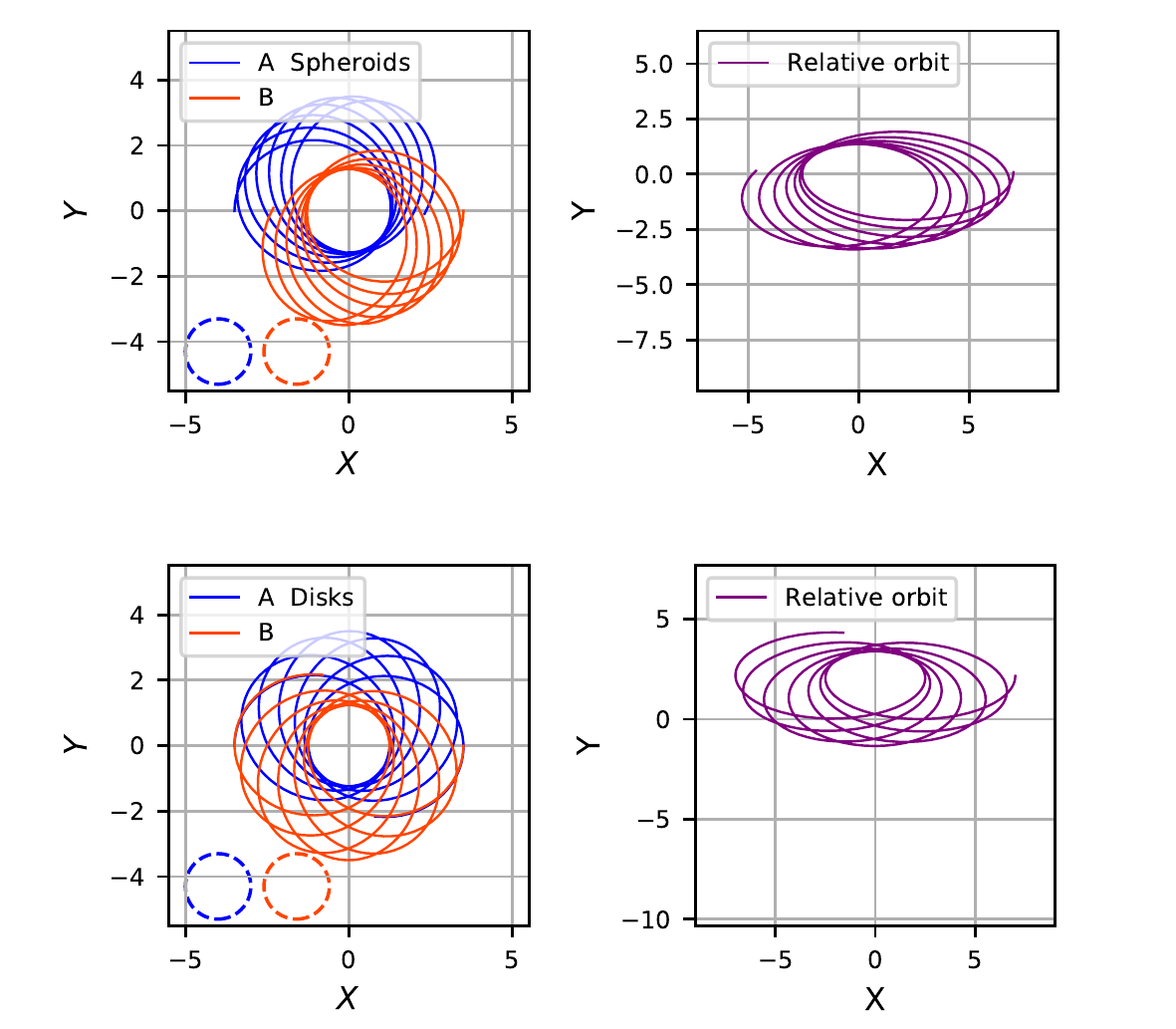}
\caption{Simulated trajectories with two identical spheroids (top row) and
  two idential disks (bottom row).
    The left-hand panels show the trajectories in the centre of mass system,
    and the right-hand panels show the relative trajectory of B
    with respect to A.
  The size of each body (drawn to scale) in the equatorial plane is indicated
  with dashed circles. The initial position and velocity
  for bodies A and B were $\mathbf{R_A} = [-3.5,0,0]$,
  $\mathbf{R_B}=[3.5,0,0]$, $\mathbf{V_A}=[0,0.2,0]$ and
  $\mathbf{V_B}=[0,-0.2,0]$. Both simulations are for $t=300$ time units.
  All quantities are unit-less}
\label{fig:fig2}
\end{figure}

We also include a model with two thin disks to test the equations in
Sect.~\ref{section:sec3}. The two disks are identical with the same masses
as the spheorids, see the 
details in Table~\ref{table:Table1}.
As the force and mutual potential are given by analytic expressions,
the two disks compute much faster
than two spheroids, hence we chose a shorter time step (set to 0.01) for
the RK integrator. The resulting
trajectories are shown in the bottom two panels of Fig.~\ref{fig:fig2}.
As can be seen, the trajectories for the two disks are very similar to
that for the two spheroids, except that the precession angle and
precession rate are larger, with $\psi _{p}=30^{\circ }$ and
$\Omega_{p}=38^{\prime }$ per time unit.

\subsection{Conservation of energy and orbital angular momentum}
\label{sec:erroranalysis}

\begin{figure*}
\includegraphics[width=1.0\textwidth]{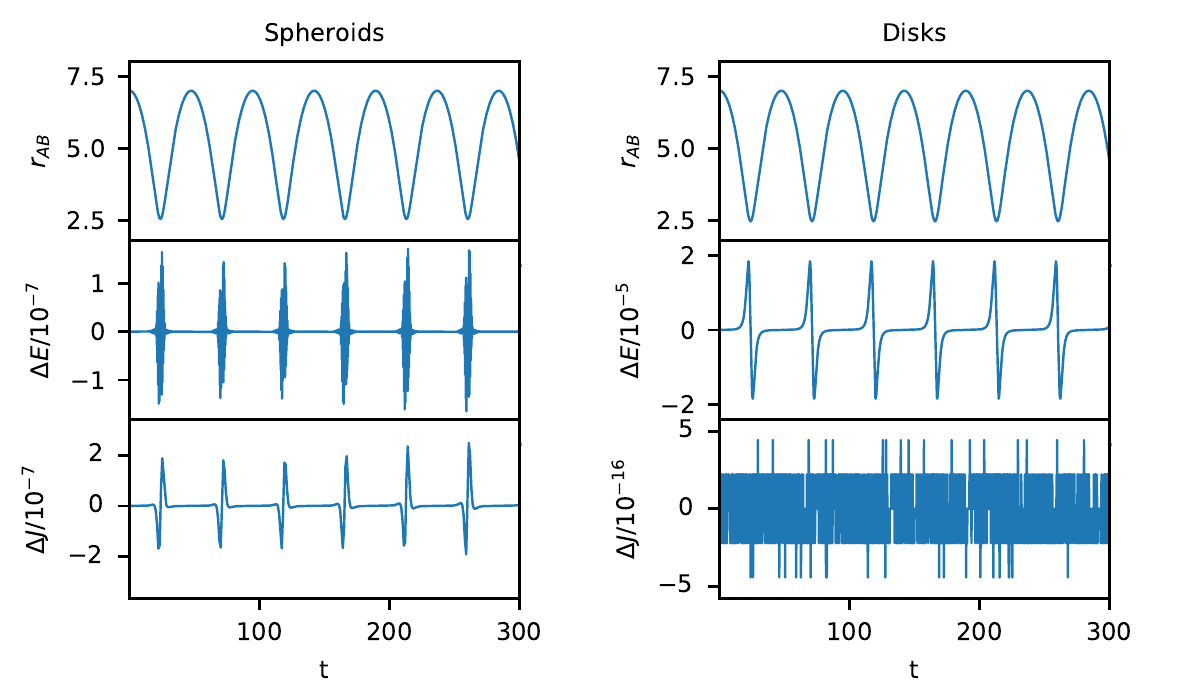}
\caption{Error analysis of the simulations with two spheroids (left) and
  two disks (right). The centroid-to-centroid distance is $r_{AB}$,
  and $\Delta E$ and $\Delta J$ are the change in energy and angular
  momentum between successive time steps. Simulation time is $t$}
\label{fig:fig3}
\end{figure*}

Because no external forces or torques are applied to the system, the total
energy and orbital angular momentum are conserved. In reality
however, these two quantities are affected both by the finite numerical
precision of the simulations (numerical errors), and by the finite
time-stepping of the integration scheme (integration errors).
In order to analyze the accuracy and energy drift
in our simulations, we probe the change in total energy and angular momentum
between successive time steps, termed $\Delta E$ and $\Delta J$. In the absence of
errors, both these quantities should be zero. The results are displayed in 
Fig.~\ref{fig:fig3}, where also a figure of how the 
distance $r_{AB}$ between the two centroids changes throughout the simulation
is included.

\begin{figure*}
\includegraphics[width=1.0\textwidth]{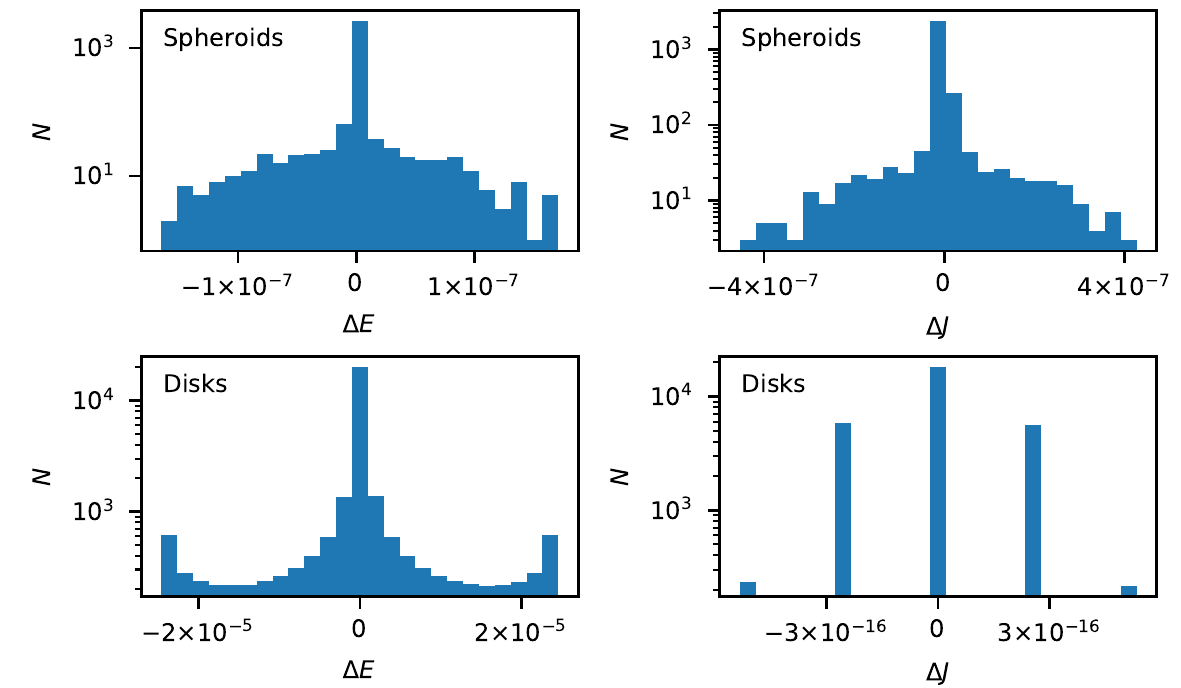}
\caption{The distributions of $\Delta E$ and $\Delta J$ in the two cases discussed in the text. 
The left-hand panels show $\Delta E$, and the right-hand panels show $\Delta J$}
\label{fig:fig4}
\end{figure*}

As expected, the errors increase as $r_{AB}$ approaches a minimum
(pericentre) where the two bodies
undergo the most rapid changes in kinetic and potential energy. The maximum
value of $\Delta E$ and $\Delta J$ for the simulation with the two spheroids is 
$1-2 \times 10^{-7}$. In other simulations (not showed here) where the
initial distance between the two bodies is three times larger, the maximum
errors decrease by 2--3 orders of magnitude. For the two disks, where an analytical
expression is used, the errors in $\Delta J$ are at the machine accuracy level.
The distributions of $\Delta E$ and $\Delta J$ are displayed as histograms
in Fig.~\ref{fig:fig4},
and the accompanying Table~\ref{table:errors} lists characteristics of these
distributions such as the mean, the standard
deviation and the standard error. We have also fitted a straight line to $\Delta E$ and $\Delta J$
as a function of time to measure if there is some drift in energy or angular momentum over
the simulation, and the slope is also listed in Table~\ref{table:errors}.

We expect both numerical errors
and integration errors in the simulation, however the integration errors are
likely small as they accumulate over time and our simulations are relatively short (just a few orbits).
The numerical errors are random in
nature and should have a mean of zero, whereas integration errors
accumulating over time should have a non-zero average  
(\citeauthor{ep16} \citeyear{ep16}). If integration errors are negligible (which we argue that they
are given the small time steps and the short duration of the simulations),
then $\Delta E$ and $\Delta J$ should be dominated by random errors and their means should be
close to zero. The standard deviations in $\Delta E$ and $\Delta J$ will in that case
reflect the magnitude of the random errors.

Fig.~\ref{fig:fig4} and the numbers in Table~\ref{table:errors} show that 
$\Delta E$ and $\Delta J$ have mean values of $<10^{-8}$, hence
integration errors are negligible. 
The standard deviations $\sigma$ are two to four orders of magnitude larger than the mean,
also suggesting that the simulations are dominated by random rather than integration errors. 
There is also negligible drift 
over the duration of the simulations as the slopes are smaller than the standard deviations. 

Table~\ref{table:errors} also lists 
the standard error, $\sigma/\sqrt{N}$, which is a measure of the
uncertainty in the mean, and we find values that are typically an order
of magnitude larger than the mean, indicating that the mean cannot be determined
properly. This happens because the changes in 
energy and angular momentum are correlated at short time scales
(smaller variations on shorter time scales superimposed on variations
happening on longer time scales).

\subsection{Limiting cases for spheroids: point masses and thin disks}
\label{section:sec3a}

As a final check, we investigate how two cases with 1) nearly spherical 
and 2) very flattened spheroids compare with two interacting point masses and two interacting thin disks.
In other words, we are comparing two cases were we use the surface integrals in Sect.~\ref{section:sec1}
with the two cases with known analytical solutions for the mutual gravitational force, i.e. that of
point masses and thin disks. We are thus testing the surface integration at the limit when it approaches
the analytical solutions. 

\begin{figure}
  \includegraphics{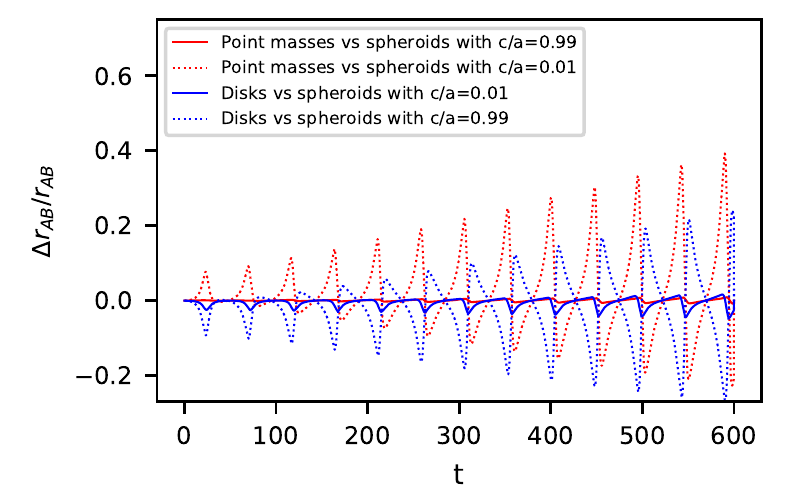}
  \caption{Comparing cases of nearly spherical ($c/a=0.99$) and significantly flattened ($c/a=0.01$) spheroids with
    point masses and thin disks. The solid lines show how nearly spherical spheroids compare to point masses (red), and
    how nearly flat spheroids compare with thin disks (blue). 
    The dotted lines compare cases that are clearly expected to be poor comparisons, i.e.\ nearly spherical spheroids
  with disks, and flattened spheroids with point masses}
\label{fig:fig7}
\end{figure}

First, we set $c/a=0.99$ for the two spheroids which make them nearly spherical, and thereafter 
$c/a=0.01$ which flattens them considerably so that they become comparable to two disks.
The mass ratio is kept at $m_A/m_B=1$
for all cases, and the systems are given identical initial conditions.
A pairwise comparison is made in Fig.~\ref{fig:fig7} where we plot the relative difference in
the distance between the two bodies, i.e.\ $\Delta r_{AB}/r_{AB}$. 
The dotted lines show the relative differences 
between point masses and flat spheroids, and between thin disks and the $c/a=0.99$
spheroids, and are of course the largest as these are cases that are not really comparable.
However, the relative difference
decreases when we compare $c/a=0.99$ spheroids with point masses, and $c/a=0.01$
spheroids with thin disks, confirming that 
the surface integration model produces results that approach the analytical cases.
Finally, we note that the relative difference peaks each time 
the bodies are close, and that the amplitude at periapsis increases over time.
From this figure, we also note that using two point masses as an approximation to
two nearly spherical
spheroids seems to be produce a better approximation than using two thin disks as an approximation
to two flattened spheroids.  

\section{Summary and conclusions}

\label{section:sec5}

We explore the application of a surface integration method to compute the force and mutual 
gravitational potential between two extended, rigid bodies. 
We first briefly describe how the surface integral expressions are obtained by vector calculus
following Conway's publication (\citeauthor{conway16} \citeyear{conway16}), and thereafter
how the surface integrals can be computed for cases of two interacting ellipsoids/spheroids. 
By assuming that the gravitational 
potential of one body can be analytically expressed as a spheroid (\citeauthor{macmillan}
\citeyear{macmillan}), and integrating over a spheroid assumed to be the second body,
we solve the equations of motion to test the method in a few simple planar cases.
The resulting trajectories are non-closed
orbits with either positive or negative precession, typical for bodies moving in non-Keplerian
potentials.

The surface integration scheme outlined in this paper can be applied to a spheroid or an ellipsoid, and if
the other body has a gravitational potential that can be expressed analytically
(like for ellipsoids and spheroids), the mutual force, torque and gravitational potential can be computed
exactly between the two bodies (to within the limits of numerical integration over the
surface of one of the bodies), thus avoiding truncation errors associated with series expansions
of potentials. In a forthcoming paper, we apply the method to two ellipsoids
with a full three dimensional treatment including torques (Ho et al., in prep.).

If the two bodies are coplanar, non-coaxial, infinitely thin disks, the surface integrals can be solved analytically,
hence we also test the analytical expressions given by
\citeauthor{conway16} (\citeyear{conway16}). Moreover, we demonstrate that 
in two limiting cases of almost spherical and significantly flattened
spheroids, the spheroid solutions approach that of the analytical solutions for 
point masses and thin disks.

We have used the analytic expression by MacMillan (\citeyear{macmillan}) for the gravitational
potential of a spheroid inside the surface integral, but is it also possible to replace this with other
forms of arbitrary potentials for studying problems involving e.g.\ an ellipsoid/spheroid and an arbitrary body.

The method we present is promising for studying dynamics between two solid,
non-spherical bodies, for instance two ellipsoids or two spheroids, or a combination of the
two. Since the method is exact and avoids using series expansions for gravitational potentials,
it is applicable, and will produce exact results, in situations where the two bodies are close to each other.
It is therefore interesting to apply it for studying dynamics of asteroids binaries, in particular
contact binaries and cases of rotational fission where the two bodies are initially in contact with
each other. 

\begin{acknowledgements}
  The authors thank the referee for helpful comments and suggestions that
  improved the presentation of the work and the manuscript.
\end{acknowledgements}

%
\section*{Conflict of interest}
The authors declare that they have no conflict of interest.

\bibliographystyle{plainnat}

\begin{thebibliography}{}
%

\bibitem[Ashenberg(2007)]{ashenberg07}
  Ashenberg, J. Mutual gravitational potential and torque of solid bodies via inertia integrals. CMDA {\bf 99}, 149--159 (2007)
   \bibitem[Binney \& Tremaine(2008)]{btbook}
   Binney, J., Tremaine, S. Galactic dynamics. Princeton University Press, New Jersey (2008)
\bibitem[Borderies(1978)]{borderies78}{}
  Borderies, N. Mutual gravitational potential of N solid bodies. Celestial Mechanics {\bf 18}, 295--307 (1978)
\bibitem[Bou{\'e} \& Laskar(2009)]{bl09}
  Bou{\'e}, G., Laskar, J. Spin axis evolution of two interacting bodies. Icarus {\bf 201}, 750--767 (2009)
\bibitem[Conway(2016)]{conway16}{}
  Conway, J. T. Vector potentials for the gravitational interaction of extended bodies and laminas with
analytical solutions for two disks. CMDA {\bf 125}, 161--194 (2016)
\bibitem[Eastman \& Pande(2016)]{ep16}
Eastman, P., Pande, V.S. Energy Conservation as a Measure of Simulation Accuracy. bioRxiv 083055 (2016)
\bibitem[Fahnestock \& Scheeres(2006)]{fs06}
     Fahnestock, E. G., Scheeres, D. J. Simulation of the full two rigid body problem using polyhedral mutual
       potential and potential derivatives approach. CMDA {\bf 96}, 317--339 (2006)
\bibitem[Fukushima(2014)]{fukushima14}
   Fukushima T. Prolate spheroidal harmonic expansion of gravitational field. AJ {\bf 147}, 152--160 (2014)
 \bibitem[Garmier \& Barriot(2001)]{gb01}
   Garmier, R., Barriot, J.-P. Ellipsoidal Harmonic expansions of the gravitational potential:
   Theory and application. CMDA {\bf 79}, 235--275 (2001)
 \bibitem[Hirabayashi \& Scheeres(2013)]{hs13}
  Hirabayashi, M., Scheeres, D.J. Recursive computation of mutual potential between two polyhedra.
  CMDA {\bf 117}, 245--262 (2013)
\bibitem[Hou, Scheeres \& Xin(2017)]{hou2017}
  Hou, X., Scheeres, D.J., Xin, X. Mutual potential between two rigid bodies with arbitrary
  shapes and mass distributions. CMDA {\bf 127}, 369--395 (2017)
\bibitem[Hartmann(1994)]{hartmann94}{}
  Hartmann, T., Soffel, M. H., Kioustelidis, T. On the use of STF-tensors in celestial mechanics.
  CMDA {\bf 60}, 139--159 (1994)
\bibitem[Jacobson \& Scheeres(2011)]{jacobsonscheeres2011}
  Jacobson, S. A., Scheeres, D. J. Dynamics of rotationally fissioned asteroids: Source of observed
  small asteroid systems. Icarus {\bf 214}, 161--178 (2011)
  \bibitem[Jekeli(1988)]{jekeli88}
    Jekeli, C. The exact transformation between ellipsoidal and spherical harmonic expansions.
    Manuscripta Geodaetica {\bf 13}, 106--113 (1988)
\bibitem[Maciejewski(1995)]{maciejewski95}{}
  Maciejewski, A. J. Reduction, Relative Equilibria and Potential in the Two Rigid Bodies Problem. CMDA {\bf 63}, 1--28 (1995)
 \bibitem[MacMillan(1930)]{macmillan}
   MacMillan W.D. Theory of the potential. McGraw-Hill, New York (1930).
\bibitem[Margot et al.(2002)]{margot02}
  Margot, J. L., Nolan, M. C., Benner, L. A. M., Ostro, S. J., Jurgens, R. F.,
  Giorgini, J. D., Slade, M. A., Campbell, D. B.:
  Binary Asteroids in the Near-Earth Object Population. Science {\bf 296}, 1445--1448 (2002).
 \bibitem[Reimond \& Baur(2016)]{rb16}
   Reimond, S., Baur, O. Spheroidal and ellipsoidal harmonic expansions of the
   gravitational potential of small Solar
   System bodies. Case study: Comet 67P/Churyumov-Gerasimenko. Journal of Geophysical
   Research (Planets) {\bf 121}, 497--515 (2016)
\bibitem[Scheeres(2009)]{scheeres09}
  Scheeres, D.~J. Stability of the planar full 2-body problem. CMDA {\bf 104}, 103--128 (2009)
\bibitem[Shi, Wang \& Xu(2017)]{shi17}
         Shi, Y., Wang, Y., Xu, S. Mutual gravitational potential, force, and torque of a homogeneous polyhedron and an extended body: an application to binary asteroids. CMDA {\bf 129}, 307--320
\bibitem[Werner \& Scheeres(2005)]{ws05}
   Werner, R.A., Scheeres, D.J. Mutual Potential of Homogeneous Polyhedra. CMDA {\bf 91}, 337--349 (2005)
\end{thebibliography}


\end{document}